\documentstyle[prl,aps,multicol,epsf]{revtex} 
\begin{document}
\draft

\title{From Massively Parallel Algorithms and Fluctuating Time Horizons to
Non-equilibrium Surface Growth}
\author{G. Korniss,$^1$ Z. Toroczkai,$^{2,3}$ M. A. Novotny,$^1$ and
P. A. Rikvold$^{1,4}$}
\address{$^1$Supercomputer Computations Research Institute, Florida State 
University, Tallahassee, Florida 32306-4130 \\
$^2$Department of Physics, University of Maryland, College Park, 
MD 20742-4111 \\
$^3$Department of Physics, Virginia Polytechnic Institute and State 
University, Blacksburg, VA 24061-0435 \\
$^4$Center for Materials Research and Technology and Department of 
 Physics, Florida State University, Tallahassee, Florida 32306-4350}

\date{January 14, 2000}
\maketitle
\begin{abstract}
We study the asymptotic scaling properties of a massively parallel algorithm 
for discrete-event simulations where the discrete events are Poisson arrivals.
The evolution of the simulated time horizon is analogous to a non-equilibrium 
surface. Monte Carlo simulations and a
coarse-grained approximation indicate that the macroscopic landscape 
in the steady state is governed by the Edwards-Wilkinson Hamiltonian.
Since the efficiency of the algorithm corresponds to the 
density of local minima in the associated surface, our results imply 
that the algorithm is asymptotically scalable.
\end{abstract}

\begin{multicols}{2}
To efficiently utilize modern supercomputers requires massively parallel 
implementations of dynamic algorithms for  various physical, chemical, and 
biological processes. 
For many of these there are well-known and routinely used serial
Monte Carlo (MC) schemes which are based on the realistic assumption that 
attempts to update the state of the system form a Poisson process.
The parallel implementation of these dynamic MC algorithms belongs
to the class of parallel discrete-event simulations, which is one of the most
challenging areas in parallel computing \cite{Fuji} and has numerous 
applications not only in the physical sciences, but also in computer science, 
queueing theory, and economics. 
For example, in lattice Ising models the discrete events are spin-flip 
attempts, while in queueing systems they are job arrivals. 
Since current special- or multi-purpose parallel computers can have 
$10^4 - 10^5$ processing elements (PE) \cite{MP}, it is essential to 
understand and estimate the scaling properties of these algorithms.

In this Letter we introduce an approach to investigate the asymptotic scaling 
properties of an extremely robust parallel scheme \cite{Luba}. This parallel
algorithm is applicable to a wide range of stochastic cellular automata with 
local dynamics, where the discrete events are Poisson arrivals.
Although attempts have been made to estimate its efficiency
under some restrictive assumptions \cite{conserv}, the mechanism which 
ensures the scalability of the algorithm in the ``steady state'' was never 
identified. Here we accomplish this by noting that the simulated time horizon 
is analogous to a growing and fluctuating surface. The {\em local} 
random time increments correspond to the deposition of random amounts of 
``material'' at the local minima of the surface. This correspondence provides 
a natural ground for cross-disciplinary application of well-known concepts 
from non-equilibrium surface growth \cite{Barabasi} and driven systems 
\cite{DDS} to a certain class of massively parallel computational schemes.
To estimate the efficiency of this algorithm one must understand
the morphology of the surface associated with 
the simulated time horizon. In particular, the efficiency of this parallel 
implementation (the fraction of the non-idling processing elements) 
exactly corresponds to the density of local minima in the 
surface model. We show that the steady-state behavior of the
macroscopic landscape is governed by the Edwards-Wilkinson (EW) Hamiltonian
\cite{EW}, implying that the density of the local minima does not 
vanish when the number of PEs goes to infinity. This ensures that the
simulated time horizon propagates with a non-zero average velocity in the
steady state. Thus the algorithm is asymptotically scalable! 
Further, based on the strong analogy between the evolution
of the simulated time horizon and the single-step surface growth model 
\cite{single_step}, we describe the asymptotic scaling 
properties of the parallel scheme. 

The difficulty of parallel discrete-event simulations is that
the discrete events (update attempts) are not synchronized by a global clock.
The traditional dynamic MC algorithms were
long believed to be {\em inherently} serial, i.e., 
in spin language, the corresponding algorithm could attempt to update only one
spin at a time. Lubachevsky nevertheless presented 
an approach for parallel simulation of these systems \cite{Luba} 
{\em without} changing the dynamics of the underlying model \cite{note}. 
Applications of his scheme include modeling of cellular communication networks 
\cite{GLNW}, ballistic particle deposition \cite{LPR}, and 
metastability and hysteresis in kinetic Ising models \cite{KNR}. 

Here we consider the case of one-dimensional systems with only
nearest-neighbor interactions (e.g., Glauber spin-flip dynamics) and
periodic boundary conditions. We restrict ourselves to the case 
where each PE carries one site (e.g., one spin) of the underlying system.
Non-zero efficiency for this algorithm implies non-zero 
efficiency for the case where a PE carries a block of sites \cite{Luba,KNR}. 
Also, the stochastic model for 
the simulated time horizon is an exact mapping only for the 
{\em synchronous} algorithm, where the main simulation cycles are executed in 
lock-step on each PE. Our goal is to show for this one-site-per-PE synchronous
algorithm (which can be regarded as the worst-case scenario) that the 
efficiency does not go to zero as the number of PEs goes to infinity. 
The basic synchronous parallel scheme \cite{Luba} is as follows. 

The size of the underlying system, and thus the number of PEs, is $L$. 
Update attempts at each site are independent Poisson processes with the same 
rate, independent of the state of the underlying system. Hence, the 
random time interval between two successive attempts is exponentially 
distributed. Without loss of generality we use time increments of mean
one (in arbitrary units).
The Poisson arrivals correspond to attempted instantaneous 
changes in the state of the site. In the parallel algorithm each PE 
generates its own {\em local simulated time} for the next update attempt, 
denoted by $\tau_i(t)$, $i=1,2,\dots,L$. Here, $t$ is the discrete index of 
the parallel steps {\em simultaneously} performed by each PE. 
Initially $\tau_i(0)=0$ for each site, and an initial configuration for the
underlying system is specified. 
The simulated time of the first update attempt is determined by 
\mbox{$\tau_i(1)=\tau_i(0)+\eta_i(0)$}, where $\{\eta_i\}$ are independent 
exponential variables. For parallel steps $t\geq 1$, each PE must compare 
its local simulated time to the  local simulated times of its neighbors. If 
$\tau_i(t)\leq\min\{\tau_{i-1}(t),\tau_{i+1}(t)\}$, the change of state of 
the site is attempted (and decided by the rules of the underlying system),
and  its local simulated time is incremented by an exponentially distributed 
random amount, $\tau_i(t+1)=\tau_i(t)+\eta_i(t)$. Otherwise, the change of 
state is not attempted and the local simulated time remains the same, 
$\tau_i(t+1)=\tau_i(t)$, i.e., the PE waits (``idles''). The comparison 
of the nearest-neighbor simulated times, and waiting if necessary, ensures that
information passed between PEs does not violate causality.
The algorithm is obviously free from deadlock, since at worst 
the PE with the absolute minimum local time can make progress.
After the initial step, the probability density
of the simulated time horizon $\{\tau_i(t)\}$ is a continuous measure,
so the probability that updates for two nearest-neighbor sites are  
attempted at the same simulated time, is of measure zero. 
When modeling the efficiency, we ignore communication times between PEs,
since they typically contribute to a scalable overhead.
Thus, the efficiency is simply the fraction of
non-idling PEs (inherent utilization). This exactly corresponds to the 
density of local minima of the simulated time horizon.

The question naturally arises: Is it possible that the fraction of non-idling
PEs goes to zero in the $L$$\rightarrow$$\infty$ limit? This would obviously 
make the algorithm unscalable and the performance of the actual implementation
poor if not disastrous.
To study this problem, we focus on the evolution of the simulated time horizon
$\{\tau_i(t)\}$, which is {\em completely independent} of the underlying model.
The above algorithmic steps can be compactly summarized as:
\begin{eqnarray}
\tau_{i}(t+1) & = &  \tau_{i}(t) \label{tau_evolution} \\ 
  & + & \Theta\left(\tau_{i-1}(t)-\tau_{i}(t)\right) 
\Theta\left(\tau_{i+1}(t)-\tau_{i}(t)\right) 
\eta_{i}(t) \;. \nonumber
\end{eqnarray}
The $\eta_i(t)$ are drawn from an exponential distribution independently
at every time $t$ and site $i$, and independent of $\{\tau_i(t)\}$. Here
$\Theta(\cdot)$ is the Heaviside step-function. This stochastic evolution 
model is very simple and easily simulated on a serial computer.
Alternatively, one can consider the evolution of the local slopes, 
$\phi_i = \tau_i -\tau_{i-1}$:
\begin{eqnarray}
\phi_{i}(t+1) - \phi_{i}(t)  & = &   \label{phi_evolution}  
\Theta\left(-\phi_{i}(t)\right)\Theta\left(\phi_{i+1}(t)\right)\eta_{i}(t) \\
& - & 
\Theta\left(-\phi_{i-1}(t)\right)\Theta\left(\phi_{i}(t)\right)\eta_{i-1}(t)
\;. \nonumber
\end{eqnarray}
The periodic boundary conditions for $\{\tau_i\}$ impose the constraint
$\sum_{i=1}^{L}\phi_i =0$.
In this representation the operator for the density of local minima is 
\begin{equation}
u(t)=\frac{1}{L}\sum_{i=1}^{L}\Theta(-\phi_{i}(t)) \Theta(\phi_{i+1}(t)) \;.
\label{density_operator}
\end{equation}
The process described by (\ref{phi_evolution}) is a microscopic
realization of biased diffusion \cite{DDS}. The 
random amount of material ``deposited'' at a local minimum $\tau_i$ 
corresponds to the transfer of this amount from $\phi_{i+1}$ to $\phi_i$. 
Since the noise in (\ref{phi_evolution}) is 
independent of $\{\phi_i(t)\}$, the average can be simply taken. This yields
a transparent continuity equation,
$\langle\phi_i (t+1)\rangle - \langle\phi_i (t)\rangle = 
-(\langle j_i\rangle - \langle j_{i-1}\rangle )$,
where the average current is 
$\langle j_i\rangle = -\langle\Theta(-\phi_i) \Theta(\phi_{i+1})\rangle$.
Translational invariance implies that
$\langle u\rangle=\langle\Theta(-\phi_i) \Theta(\phi_{i+1})\rangle$, 
which is the same as the magnitude of the average current or the mean velocity
of the surface.

To gain some insight into the evolution of the surface,
we perform a naive coarse-graining by taking an ensemble average on 
(\ref{phi_evolution}) and replacing $\Theta(\phi)$ with a smooth 
representation.
The procedure is independent of the actual form of the representation.
We use $\Theta_{\kappa}(\phi) = (1/2)[\tanh(\phi/ \kappa)+1]$
so $\lim_{\kappa \rightarrow 0} \Theta_{\kappa}(\phi) = 
\Theta(\phi)$. To leading order in $\phi/ \kappa$,
\begin{eqnarray} \label{discrete_burgers}
\langle\phi_i (t+1)\rangle - \langle\phi_i (t)\rangle & = & 
\frac{1}{4\kappa} \langle \phi_{i+1} -2\phi_{i} + \phi_{i-1} \rangle \\
& - & \frac{1}{4\kappa^2} \langle \phi_{i}(\phi_{i+1} -\phi_{i-1}) \rangle
\;.\nonumber
\end{eqnarray}
Taking the naive continuum limit 
one obtains 
\begin{equation}
\partial_t \hat{\phi} = 
\frac{\partial^2 \hat{\phi}}{\partial x^2} - 
\lambda \frac{\partial}{\partial x} \hat{\phi}^2
\label{continuum_burgers}
\end{equation}
for the coarse-grained field, $\hat{\phi}$,
where roughly speaking $\lambda$, the coefficient of the nonlinear term, 
carries the details of the coarse-graining procedure. Equation 
(\ref{continuum_burgers})
is the nonlinear biased diffusion or Burgers' equation \cite{Burgers}. 
Through $\hat{\phi}=\partial \hat{\tau}/\partial x$ it is simply
related to the deterministic part of the KPZ equation for the coarse-grained
surface height fluctuations \cite{KPZ},
\begin{equation}
\partial_t \hat{\tau} = 
\frac{\partial^2 \hat{\tau}}{\partial x^2} - 
\lambda \left(\frac{\partial \hat{\tau}}{\partial x}\right)^2 \;.
\label{continuum_tau}
\end{equation}
To capture the fluctuations one typically extends the 
above equations with appropriate noise, i.e., conserved for
(\ref{continuum_burgers}), and non-conserved for (\ref{continuum_tau}).
This implies that the evolution of the simulated
time horizon is KPZ-like. In one dimension the steady state of
such systems (on coarse-grained length scales) is governed by the EW 
Hamiltonian \cite{EW}, 
${\cal H}_{\rm EW}$$\propto$$\int\!\!dx (\partial \hat{\tau}/\partial x)^2$. 
This corresponds to a simple random-walk surface, where the coarse-grained 
slopes are independent in the thermodynamic limit, yielding $1/4$ for the 
density of local minima. 
Obviously this value will be different for our specific microscopic model. 
However it {\em cannot} vanish: a zero density of local 
minima in the $L$$\rightarrow$$\infty$ limit would imply that it is zero at all
length scales. This would contradict our finding that the steady 
state at the coarse-grained level is governed by the EW Hamiltonian. 
The non-zero density of local minima is an important characteristic 
of this (steady-state) universality class. It ensures that our specific 
microscopic surface propagates with a non-zero average velocity in the steady 
state. Models belonging to other universality 
classes do not necessarily have non-zero extremal-point densities. For example,
we can show \cite{toro} that the density of local minima vanishes for
a one-dimensional curvature-driven random Gaussian surface.

We now present our MC results to test the coarse-graining approach.  
First we follow the time evolution of the width
$\langle w^2(t)\rangle$$=$$(1/L) \langle
\sum_{i=1}^{L}\left[\tau_i(t)-\overline\tau(t)\right]^2 
\rangle$, where $\overline\tau(t)$$=$$(1/L)\sum_{i=1}^{L}\tau_i(t)$ 
and the average $\langle\cdot\rangle$ is taken over many independent runs.
After the early-time regime, which is strongly affected by the 
intrinsic width, and before saturation, we find 
$\langle w^2(t)\rangle$$\sim$$t^{2\beta}$ [Fig.\ 1(a)]. Although the system 
exhibits very strong corrections to scaling, for our largest system, 
$L$$=$$10^5$, we find $\beta$$=$$0.326\pm0.005$, which includes within two 
standard errors the exact KPZ exponent, $1/3$.
In the steady state the width is stationary and 
$\langle w^2\rangle$$\sim$$L^{2\alpha}$ for large $L$. Here the corrections 
to scaling are somewhat smaller than in the earlier regime, and the 
above scaling is obeyed for $L$$\geq$$10^3$ with $\alpha$$=$$0.49\pm0.01$. 
This agrees with the prediction that the long-distance behavior is governed by 
the EW Hamiltonian, for which $\alpha$$=$$1/2$. 
Further, plotting rescaled width $\langle w^2\rangle/L^{2\alpha}$ vs 
rescaled time $t/L^z$, with $z$$=$$\alpha/\beta$, confirms dynamic scaling 
for the intermediate-to-late time crossover \cite{FV} [Fig.\ 1(a) inset]. 
We also measured the average steady-state structure factor of $\{\tau_i\}$, 
finding the expected $\sim 1/k^2$ behavior for small wave vectors $k$. 
The spatial two-point correlations decay linearly for $\{\tau_i\}$ and are
short ranged for the slopes, $\{\phi_i\}$. 

To further probe the universal properties of the surface in the steady state
we construct the {\em full} width distribution, $P(w^2)$.
The EW class is characterized by a universal scaling function, $\Phi(x)$, such
that $P(w^2)=\langle w^2\rangle^{-1}\Phi(w^2/\langle w^2\rangle)$ \cite{FORWZ},
where 
\begin{equation}
\Phi(x)=\frac{\pi^2}{3}\sum_{n=1}^{\infty}(-1)^{n-1} n^{2} 
e^{ -\frac{\pi^2}{6} n^{2} x } \;.
\label{exact_hist}
\end{equation}
Systems with $L$$\geq$$10^3$ show convincing data collapse [Fig.\ 1(b)] onto 
this exact scaling function.  
\begin{figure}[t]
\vspace*{-1.7cm}
\hspace*{-0.7cm}
\epsfxsize=9.0cm\epsfysize=9.0cm\epsfbox{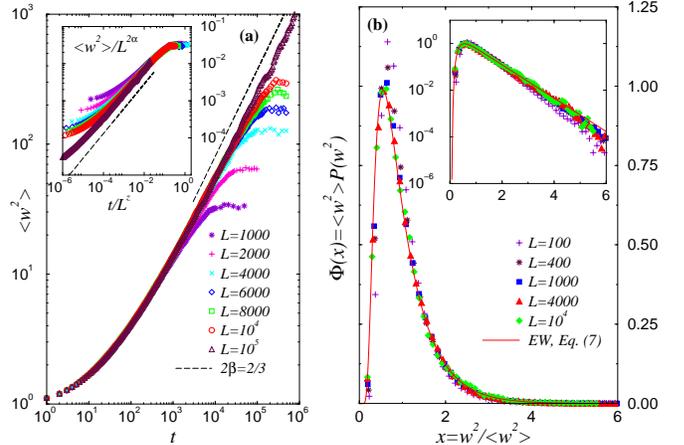}
\vspace*{-1.8cm}
\caption{(a) Time evolution and dynamic scaling (inset) for the surface 
width. (b) Steady-state width distribution (inset: on linear-log scales).}
\label{fig1}
\end{figure}

Next we estimate the efficiency of the algorithm.
When simulating the system described by (\ref{tau_evolution}) and measuring the
average local minimum density $\langle u(t)\rangle$, we observe that for 
every system size it monotonically decreases as a function of time and
approaches a {\em constant} slightly smaller than $1/4$ for large systems. 
Using the close similarity between our model and the single-step 
solid-on-solid surface-growth model \cite{single_step}, we can 
understand the scaling behavior for the {\em steady-state} average 
$\langle u \rangle$ and the fluctuations 
$\sigma^2 = \langle u^2 \rangle -\langle u \rangle^2$. In the single-step
model the height 
differences (i.e., the local slopes) are restricted to $\pm 1$, and the 
evolution consists of particles of height $2$ being deposited at the local
minima. The advantage of this model is that
it can be mapped onto a hard-core lattice gas for which the steady-state
probability distribution of the configurations is known exactly 
\cite{single_step,Spitzer}. This enables one to
find arbitrary moments of the local minimum density operator, analogous to
(\ref{density_operator}). For the single-step model we find that
$\langle u \rangle_L=(1/4)(1-1/L)^{-1} = 1/4 + 1/(4L) + {\cal O}(1/L^2)$, and
$\sigma_{L}^{2} = 1/(16L) + {\cal O}(1/L^2)$.
We propose that the scaling of the local minimum density for large $L$ in our 
model follows the same  form, i.e.,
\begin{equation}
\langle u \rangle_L - \langle u \rangle_{\infty} \propto 
L^{-1} \;, \;\;\; 
\sigma_{L} \propto L^{-1/2}
\label{params}\;.
\end{equation}
Our reasons are: 
({\it i}) both models in  their steady states belong to the EW universality 
class (short-range correlated local slopes), which guarantees that 
$\langle u \rangle_{\infty}$ is non-zero; ({\it ii}) the constraint 
$\sum_{i=1}^{L}\phi_i=0$ in our model and the conservation of the 
number of particles in the lattice gas will produce similar 
finite-size effects. Our simulation results show very good agreement with 
(\ref{params}) [Fig.\ 2(a)].
Further, extrapolating to $L$$\rightarrow$$\infty$, yields 
$\langle u \rangle_{\infty}$$=$$0.24641\pm(7\times 10^{-6})$. 
The scaling relation (\ref{params}) 
also implies that $u$ is a self-averaging macroscopic quantity: its
full distribution $P_{L}(u)$, for large $L$, is
Gausian with the parameters in (\ref{params}) [Fig.\ 2(b)]. Thus, in the 
$L$$\rightarrow$$\infty$ limit it approaches a delta-function centered about 
$\langle u \rangle_{\infty}$. 
\begin{figure}[t]
\vspace*{-2.0cm}
\hspace*{-0.6cm}
\epsfxsize=9.0cm\epsfysize=9.0cm\epsfbox{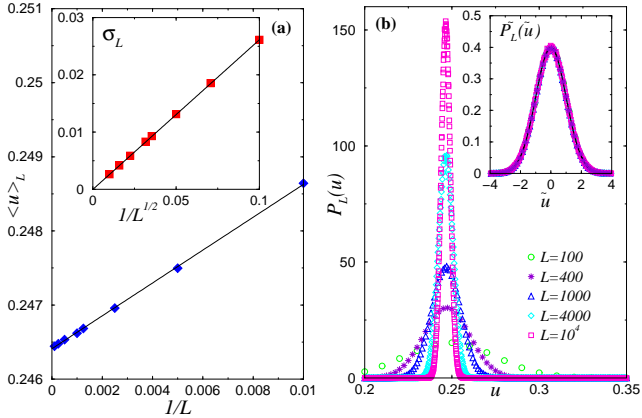}
\vspace*{-1.9cm}
\caption{Steady-state scaling for the parallel efficiency (density of local 
minima). (a) The average and the variance (inset). (b) The full probability 
density. Inset: scaled probability densities, $\tilde{P}_{L}(\tilde{u})$ with
$\tilde{u} = (u -\langle u \rangle_L)/\sigma_{L}$, collapse onto a Gaussian 
with zero mean and unit variance.}
\label{fig2}
\end{figure}

Finally, we point out the lack of up-down symmetry in our model
in the steady-state. This is most easily noticeable at short distances, 
either by looking at a snapshot [Fig.\ 3(a)], or through the high degree of 
asymmetry in the nearest-neighbor two-slope distribution [Fig.\ 3(b)]: 
the hilltops are sharp and the valley-bottoms are flattened. 
Such stationary-state skewness is generally observable in one-dimensional 
KPZ growth, but has only recently received serious attention 
\cite{KPZ_skewness,miro}.
\begin{figure}[t]
\vspace*{-4.5cm}
\hspace*{-1.5cm}
\epsfxsize=11.0cm\epsfysize=11.0cm\epsfbox{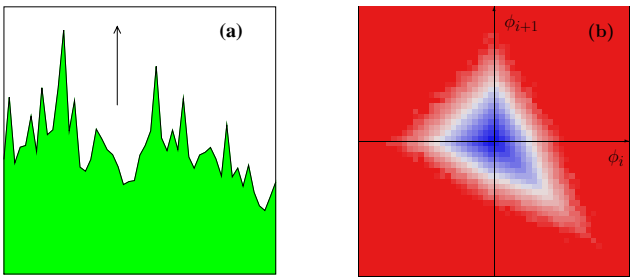}
\vspace*{-3.0cm}
\caption{(a) A short segment (50 sites) of a typical steady-state surface 
configuration. (b) Density plots for the nearest-neighbor two-slope 
distribution, $L$$=$$10^{4}$.} 
\label{fig3}
\end{figure}

In summary, we studied the asymptotic scaling properties of a
general parallel algorithm by regarding the simulated time horizon as a 
non-equilibrium surface. We conclude that the basic algorithm
(one site per PE) is scalable for one-dimensional arrays. The same 
correspondence can be applied to model the performance of the algorithm for
higher-dimensional logical PE topologies. While this will involve the typical
difficulties of surface-growth modeling, such as an absence of exact results 
and very long simulation times, it establishes potentially fruitful 
connections between two traditionally separate research areas.  

We thank S.\ Das Sarma, S.\ J.\ Mitchell, and G.\ Brown for stimulating 
discussion. We acknowledge the support of DOE through SCRI-FSU, NSF-MRSEC at 
UMD, and NSF through Grant No. DMR-9871455.

\end{multicols}


\vspace*{-0.3cm}
\begin{references}
\vspace*{-1.4cm}
\bibitem{Fuji} R.\ M.\ Fujimoto, Commun. ACM {\bf 33}, 30 (1990).

\bibitem{MP} For example the $9472$-node ASCI Red at Sandia, the   
$12,288$-node QCDSP Teraflop Machine at Brookhaven, and 
the Connection Machine CM-2 with 65,536 PEs.

\bibitem{Luba} B.\ D.\ Lubachevsky, 
Complex Systems {\bf 1}, 1099 (1987); 
J. Comput. Phys. {\bf 75}, 103 (1988).

\bibitem{conserv} B.\ D.\ Lubachevsky, in {\it Distributed Simulation 1989},
SCS simulation series, 1989 vol.\ 21, p. 100; 
D. M. Nicol, J. ACM {\bf 40}, 304 (1993).

\bibitem{Barabasi} A.-L.\ Barab\'asi and H.\ E.\ Stanley, 
{\it Fractal Concepts in Surface Growth} 
(Cambridge University Press, Cambridge, 1995).

\bibitem{DDS} B.\ Schmittmann and R.\ K.\ P.\ Zia, in 
{\it Phase Transitions and Critical Phenomena} Vol.\ 17., 
edited by C. Domb and J. L. Lebowitz (Academic Press, New York, 1995).

\bibitem{EW} S.\ F.\ Edwards and D.\ R.\ Wilkinson, Proc. R. Soc. London, 
Ser A {\bf 381}, 17 (1982).

\bibitem{single_step} P.\ Meakin, P.\ Ramanlal, L.\ M.\ Sander, and 
R.\ C.\ Ball, Phys. Rev. A {\bf 34}, 5091 (1986);  
M.\ Plischke, Z.\ R\'acz, and D.\ Liu, Phys. Rev. B {\bf 35}, 3485 (1987). 

\bibitem{note}
This contrasts with other algorithms \cite{multispin,cluster} that obtain the 
correct equilibrium properties but change the dynamics.

\bibitem{multispin} R.\ Friedberg and J.\ E.\ Cameron, J. Chem. Phys. 
{\bf 52} 6049 (1970).

\bibitem{cluster} R.\ H.\ Swendsen and J.-S.\ Wang, Phys. Rev. Lett. 
{\bf 58}, 86 (1987);
Y.\ S.\ Choi, J.\ Machta, P.\ Tamayo, L.\ X.\ Chayes, Int. J. Mod. Phys. C 
{\bf 10}, 1 (1999) and references therein. 

\bibitem{GLNW} A.\ G.\ Greenberg, B.\ D.\ Lubachevsky, D.\ M.\ Nicol, and 
P. E. Wright,
{\em Proceedings, 8th Workshop on Parallel and Distributed 
Simulation (PADS '94)}, Edinburgh, UK, (1994) p.\ 187.

\bibitem{LPR} B.\ D.\ Lubachevsky, V.\ Privman, and S.\ C.\ Roy,
J. Comput. Phys. {\bf 126}, 152 (1996).

\bibitem{KNR} 
G.\ Korniss, M.\ A.\ Novotny, and P.\ A.\ Rikvold, 
J. Comput. Phys., {\bf 153}, 488 (1999). 

\bibitem{Burgers} M.\ Burgers, {\it The Nonlinear Diffusion Equation}, 
(Riedel, Boston, 1974).

\bibitem{KPZ} M.\ Kardar, G.\ Parisi, and Y.-C.\ Zhang, Phys. Rev. Lett. 
{\bf 56}, 889 (1986). 

\bibitem{toro} Z.\ Toroczkai, G.\ Korniss, S.\ Das Sarma, and R.\ K.\ P.\ Zia, 
to be submitted.

\bibitem{FV} F.\ Family and T.\ Vicsek, J.\ Phys. A {\bf 18} L75 (1985).

\bibitem{FORWZ} G.\ Foltin, K.\ Oerding, Z.\ R\'acz, R.\ L.\ Workman, 
and R. K. P. Zia, Phys. Rev. E {\bf 50}, R639 (1994). 

\bibitem{Spitzer} F.\ Spitzer, Adv. Math. {\bf 5}, 246 (1970).

\bibitem{KPZ_skewness} J. Neergaard and M. den Nijs,
J. Phys. A {\bf 30}, 1935 (1997).

\bibitem{miro} P.\ A. Rikvold and M. Kolesik, e-print cond-mat/9909188 (1999).

\end{references}
\end{document}